\documentclass[sigconf,nonacm=true]{acmart}

\usepackage{pdfpages}
\usepackage{amsfonts}
\usepackage{pifont}
\usepackage{colortbl}
\usepackage{xcolor}
\usepackage{tcolorbox}
\usepackage{microtype}
\usepackage{multirow, booktabs}
\usepackage{adjustbox}
\usepackage{enumitem}
\usepackage{float}
\usepackage{bm}
\usepackage{bbm} 
\usepackage{amsmath}
\usepackage{graphicx}
\usepackage{hyperref}
\usepackage{soul}
\usepackage[labelformat=simple]{subcaption}
\usepackage{color}
\usepackage{fancyhdr}

\newcommand{\etal}{\textit{et al.}}

\begin{document}
\title{A Preliminary Study of ChatGPT on News Recommendation: Personalization, Provider Fairness, Fake News}

\author{Xinyi Li}
\affiliation{%
\institution{Northwestern University, IL, US}
\city{}
\state{}
\country{}
}
\email{xinyili2024@u.northwestern.edu}

\author{Yongfeng Zhang}
\affiliation{%
\institution{Rutgers University, NJ, US}
\city{}
\state{}
\country{}
}
\email{yongfeng.zhang@rutgers.edu}

\author{Edward C. Malthouse}
\affiliation{%
\institution{Northwestern University, IL, US}
\city{}
\state{}
\country{}
}
\email{ecm@northwestern.edu}
\begin{abstract}
Online news platforms commonly employ personalized news recommendation methods to assist users in discovering interesting articles, and many previous works have utilized language model techniques to capture user interests and understand news content. With the emergence of large language models like GPT-3 and T-5, a new recommendation paradigm has emerged, leveraging pre-trained language models for making recommendations. ChatGPT, with its user-friendly interface and growing popularity, has become a prominent choice for text-based tasks. Considering the growing reliance on ChatGPT for language tasks, the importance of news recommendation in addressing social issues, and the trend of using language models in recommendations, this study conducts an initial investigation of ChatGPT's performance in news recommendations, focusing on three perspectives: personalized news recommendation, news provider fairness, and fake news detection. ChatGPT has the limitation that its output is sensitive to the input phrasing. We therefore aim to explore the constraints present in the generated responses of ChatGPT for each perspective. Additionally, we investigate whether specific prompt formats can alleviate these constraints or if these limitations require further attention from researchers in the future. We also surpass fixed evaluations by developing a webpage to monitor ChatGPT's performance on weekly basis on the tasks and prompts we investigated. Our aim is to contribute to and encourage more researchers to engage in the study of enhancing news recommendation performance through the utilization of large language models such as ChatGPT.

\end{abstract}

\ccsdesc[500]{Information systems~Recommender systems}
\keywords{ChatGPT, Large-language Models, News Recommendations}

\maketitle

\section{Introduction}

In today's information-overloaded society, online platforms like Google News and Microsoft News are attracting users to read news online \cite{wu2020mind}. However, the daily volume of new news articles poses a challenge for users to find ones that align with their interests \cite{lian2018towards}. To address this, news recommendation systems (RS) are crucial for assisting users in discovering relevant articles. News articles contain rich textual information, making language model techniques like Gated Recurrent Unit (GRU) \cite{cho2014learning}, Long-Short Term Memory (LSTM) \cite{staudemeyer2019understanding}, Convolutional Neural Networks (CNNs) \cite{chen2015convolutional}, and attention mechanisms \cite{vaswani2017attention} popular choices for modeling users' interests and comprehending article content \cite{an2019neural, wu2022news, wu2019neural}. Furthermore, pre-trained language models and prompt learning techniques have demonstrated their effectiveness in various language tasks \cite{jin2021good}, leading RS researchers to approach recommendation as a language task to leverage the power of these techniques \cite{zhang2021language, cui2022m6, geng2022recommendation}.

This study aims to evaluate ChatGPT, a prominent language model developed by OpenAI, in the context of news RS tasks. Given the success of ChatGPT in various natural language processing (NLP) tasks and the growing recognition of recommendation as a language-related task, our research focuses on three key perspectives: personalized news recommendation, news provider fairness, and fake news detection. Within each perspective, our objective is to identify limitations in ChatGPT's response generation and explore the potential effectiveness of specific prompt formats or requirements to address these limitations. Additionally, we aim to shed light on areas that might require further attention from future researchers, as certain limitations may not be easily resolved through prompt design alone. We anticipate that ChatGPT will improve and address certain concerns through user feedback. Therefore, we have developed a webpage\footnote{https://imrecommender.github.io/ChatNews/} to track its progress on the tasks we have been exploring, with updates provided on a weekly basis. We hope our study would inspire OpenAI researchers and the wider scientific community to delve deeper into improving the performance of language models such as ChatGPT in news RS tasks.

\section{Related Work}\label{section:literature}
\textbf{News Recommendation.} Existing news RS methods utilize NLP techniques like denoising auto-encoders \cite{okura2017embedding}, GRU networks and CNNs \cite{an2019neural}, and attention mechanisms \cite{wu2019neural3} to understand news content and model users' interests based on their reading behavior \cite{wu2022news, wu2019neural}. While content understanding and personalized recommendations are essential, it is equally important to address social issues associated with news RS, including filter bubbles \cite{nguyen2014exploring}, echo chambers \cite{cinelli2021echo}, the spread of fake news \cite{vosoughi2018spread}, popularity bias \cite{abdollahpouri2021user}, user-side fairness \cite{li2021towards, wu2021fairness}, and provider-side fairness \cite{qi2022profairrec, burke2018balanced, sonboli2020opportunistic}. In this study, we not only evaluate ChatGPT's zero-shot performance in personalized recommendation task but also examine whether it appropriately addresses provider bias and fake news concerns. By investigating these aspects, we aim to shed light on the broader societal implications of employing ChatGPT for news RS.

\textbf{Pre-trained Language Models and RS.} Pre-trained language models like BERT \cite{devlin2018bert} and GPT \cite{radford2018improving}, which are trained on large-scale datasets, have shown adaptability to various downstream tasks, and prompt learning techniques \cite{cho2014learning} have further improved their performance. This success has led to a shift in RS, treating recommendation tasks as language tasks. Researchers have proposed various approaches, such as converting item-based recommendation to text-based tasks and utilizing textual descriptions for user behavior \cite{zhang2021language}, employing personalized prompt learning for explainable recommendation \cite{li2022personalized}, transforming user behavior into text-based inquiries \cite{cui2022m6}, and adopting flexible text-to-text approaches for RS \cite{geng2022recommendation}. In this work, we investigate ChatGPT's zero-shot performance on news recommendation tasks, leveraging its capabilities as a pre-trained language model.

\textbf{ChatGPT.} ChatGPT has gained immense popularity within a short period leading to numerous studies that explore its strengths and limitations. Qin \etal\ \cite{qin2023chatgpt} assess ChatGPT's performance on various NLP tasks, while Bang \etal\ \cite{bang2023multitask} provide a comprehensive technical evaluation of its capabilities in multitasking, multimodal, and multilingual applications. Zhou \etal\ \cite{zhuo2023exploring} explore ethical concerns associated with ChatGPT usage. Liu \etal\ \cite{liu2023chatgpt} construct a benchmark to evaluate ChatGPT's performance in RS tasks like rating prediction, sequential recommendation, direct recommendation, explanation generation and review summarization. While ChatGPT is known to have limitations, including bias and the potential for generating fake information \cite{ray2023chatgpt}, our research aims to explore the social issues related to using ChatGPT for news recommendation, particularly provider bias and fake news detection. We investigate potential prompt formats that can help mitigate these issues or highlight areas requiring further attention. 

\section{Evaluations of ChatGPT}\label{section:experiment}
This section evaluates ChatGPT's performance in news recommendations using zero-shot approaches. We specifically focus on three key tasks: personalized recommendations, fairness of news providers, and trustworthiness of the generated responses. Our approach involves first identifying any limitations in ChatGPT's responses using simple prompts. We then construct additional prompts to address these limitations or emphasize the need for further attention to these specific issues when utilizing language models like ChatGPT for news recommendation. To facilitate reproducibility, we have made the prompts and codes available on a GitHub repository\footnote{https://github.com/imrecommender/ChatGPT-News}. For our analysis, we utilize data samples from the Microsoft News Dataset (MIND) \cite{wu2020mind}. 

\subsection{Personalized Recommendation of ChatGPT}
This subsection uses a random sample of 30 users from the MIND dataset to detect limitations and gain insights into ChatGPT's performance when it generates recommendations for individual users based on a set of unread articles.

\begin{figure*}
\centering
\includegraphics[width=\textwidth]{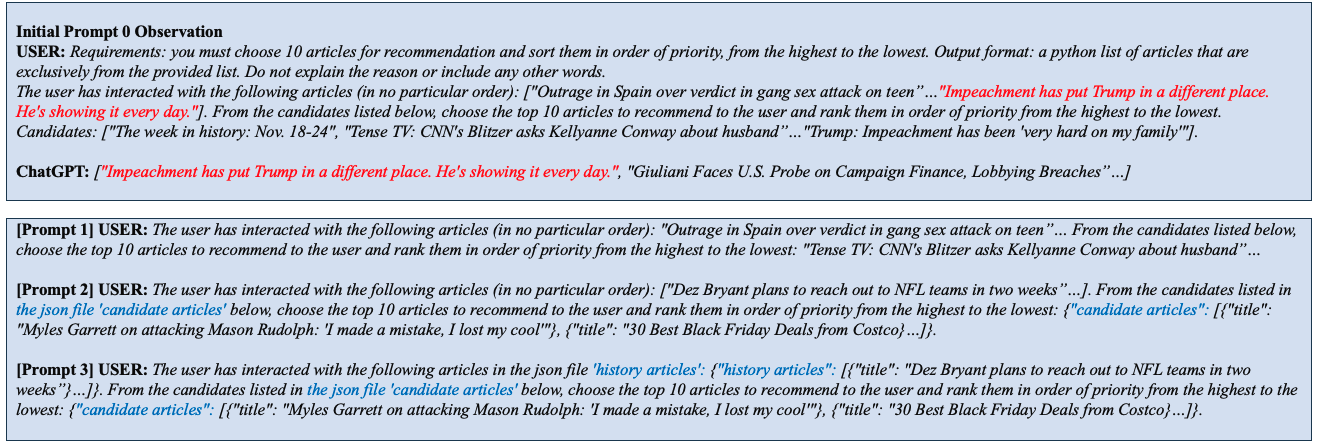}
\caption{Brief descriptions of prompts used for evaluating personalized recommendation of ChatGPT -- hypothesis 1. Using prompt 3, the proportion of ChatGPT's response containing articles read by a user is zero, with a 95\% confidence level via exact binomial test.}
\label{fig:hypothesis1}
\end{figure*} 

Based on our investigation of ChatGPT's response generation using the initial prompt provided by Liu \etal \cite{liu2023chatgpt}, we observe a limitation wherein ChatGPT struggles to effectively differentiate between articles previously read by a user and candidate articles. As a result, ChatGPT may generate recommendations that include articles already read by the user. Building upon this identified limitation, we propose the hypothesis 1:
\vspace{-0.1cm}
\begin{tcolorbox}[colback=white,colframe=black]
    \textbf{Hypothesis 1:} Improving the organization of prompts by using the JSON format with explicit keys instead of solely relying on textual descriptions will better distinguish the articles read by a user and candidate articles.
\end{tcolorbox}

We evaluate the four different prompts shown in Figure \ref{fig:hypothesis1}. We feed each prompt to the model five times for each user and count the number of users whose responses contain articles that the user has previously read. We conduct an exact binomial test to further investigate. The results indicate that when utilizing prompt 3 from Figure \ref{fig:hypothesis1}, the probability of having articles previously read by the user in the response was found to be zero. 
However, we could not reach the same conclusion for the other prompts. Based on these findings, we can infer that when dealing with lengthy texts and when it is crucial to differentiate specific information, utilizing a JSON format with explicit keys proves to be more effective than relying solely on textual descriptions. 

We further assess ChatGPT's zero-shot personalized RS capability by comparing it to several baselines, including LSTUR \cite{an2019neural}, TANR \cite{wu2019neural4} NRMS \cite{wu2019neural2}, and NAML \cite{wu2022news} using metrics top-$k$ Hit Ratio (Hit@$k$) and Normalized Discounted Cumulative Gain (nDCG@$k$). The results, presented in Table \ref{tab:hypothesis2-1}, indicate that ChatGPT's zero-shot news RS performance is inferior to existing deep neural-based models. However, we observe that there is a high probability (over 93.3\%) that the top-5 recommended articles by ChatGPT are from the same historical topics as the user's interests, whereas in the ground truth, there is only a 60\% chance that the clicked article belongs to the same categories as the historical articles. This suggests that ChatGPT is capable of understanding the categories of historical articles that users are interested in. However, user interests are dynamic, and without fine-tuning or training on the news dataset, ChatGPT's RS performance is inferior compared to existing deep neural-based models. This highlights the need for further research and potential fine-tuning approaches to enhance ChatGPT's recommendation performance in the domain of news.

\begin{table}
\centering
\small
\begin{tabular}{c|c|c|c|c}
\hline
\textbf{Model} & \textbf{Hit@5} & \textbf{nDCG@5} & \textbf{Hit@10} & \textbf{nDCG@10} \\
\hline
LSTUR & 0.5667 & 0.3674 & 0.9000 &  0.5085\\
TANR & 0.6333 & 0.3787 & 0.9333 &  0.4834 \\ 
NAML &  0.7667  &  0.4328 & 0.9333 &  0.5041\\ 
NRMS & 0.6667 &  0.4370 &  0.9333 &  0.5282\\ 
ChatGPT [prompt 3] & 0.3833 & 0.1765 & 0.7444& 0.3074\\ 
\hline
\end{tabular}
\caption{ChatGPT's zero-shot performance on personalized news recommendation, compared to baselines. }
\label{tab:hypothesis2-1}
\end{table}

\subsection{News Provider Fairness}
Most news organizations that create content (i.e., \emph{providers}) depend on advertising for a substantial fraction of their operating revenues, supplementing other revenue sources such as user-subscriber fees, cable TV carriage fees, and donations. Digital advertising depends on attracting users to the news site, and an important referring source of visitors is news, social media and search platforms, which implement RS. Reduced levels of ad revenue have contributed to news organizations closing, which has created vast news deserts in the US, where communities no longer have news coverage \cite{abernathy2018expanding}. When Facebook changed its RS in 2018 small news organizations had decreases in traffic and ad revenue \cite{Kathleen_2018}, and countries such as Australia are attempting to regulate platforms and have them pay news organizations for their content. Platforms that implement news RS must therefore balance the needs of different stakeholders with multiple objectives, and they may want to guarantee that various providers receive some ``fair'' proportion of recommendations. While provider fairness is often addressed as a post-processing in news RS \cite{burke2018balanced, wu2021tfrom}, our objective is to first identify any biases related to news provider fairness using ChatGPT and then explore potential prompt improvement to alleviate these concerns. We divide providers into two groups, popular and unpopular, and we utilize precision@$k$ to assess the proportion of popular providers among the top-$k$ recommendations.

\begin{figure*}
\centering
\includegraphics[width=\textwidth]{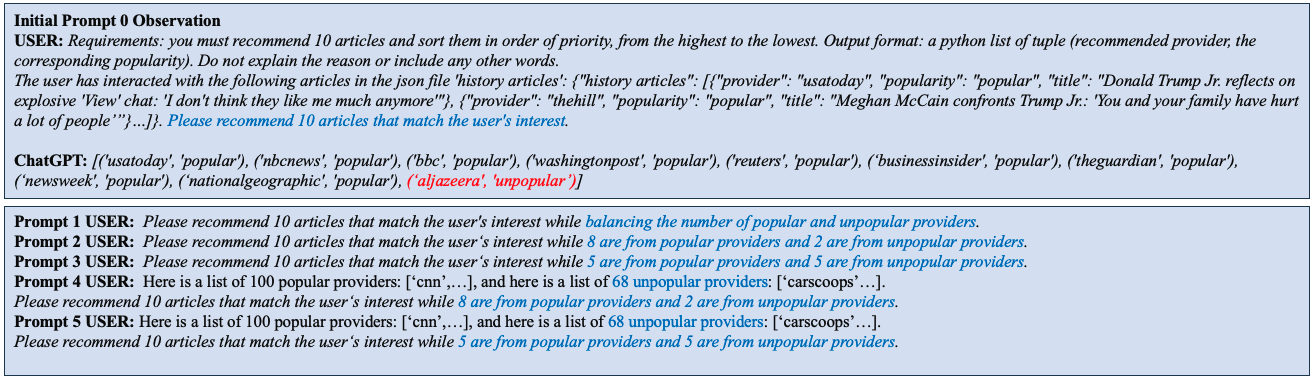}
\caption{Brief descriptions of prompts for evaluating the group-level provider fairness with no candidate article -- hypothesis 2.}
\label{fig:hypothesis2}
\end{figure*} 

The first scenario involves not providing candidate articles to ChatGPT but instead asking it for recommendations based on the articles that a user has read before. In our preliminary experiment using initial prompt 0 from Figure \ref{fig:hypothesis2}, we observe that ChatGPT mistakenly labels some popular providers as unpopular in its responses. This prompts us to further investigate provider fairness metrics from two perspectives: the user's perspective where we adjust the popularity labels based on a pre-defined list of 100 popular providers, and ChatGPT's perspective where we evaluate its performance using the popularity labels assigned by ChatGPT in its responses. Additionally, in the initial experiment, we notice that ChatGPT tends to recommend articles from providers labeled as popular by ChatGPT. This finding prompt us to propose the following hypothesis:

\vspace{-0.2cm} 
\begin{tcolorbox}[colback=white,colframe=black]
    \textbf{Hypothesis 2:} Explicitly specifying the number of articles from both popular and unpopular providers will mitigate the issue of provider bias based on a user's tolerance.
\end{tcolorbox}

\begin{figure}
    \centering
    \begin{subfigure}[b]{0.43\textwidth}
        \centering
        \includegraphics[width=\textwidth]{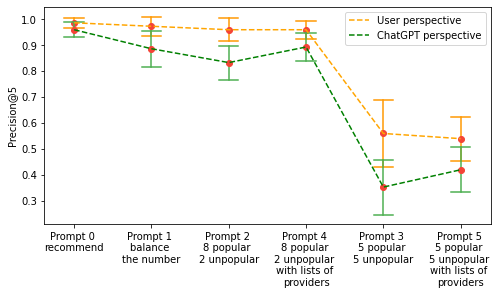}
        \label{fig:subfig1}
    \end{subfigure}
    \begin{subfigure}[b]{0.43\textwidth}
        \centering
        \includegraphics[width=\textwidth]{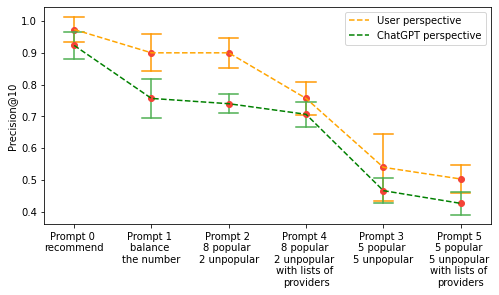}
        \label{fig:subfig2}
    \end{subfigure}
    \caption{Performance evaluation from both user and ChatGPT standpoints for provider fairness when there is no candidate provided -- hypothesis 2. The statistical t-test confirms that ChatGPT is controllable for improving the provider fairness based on users' tolerance.}
     \label{fig:generative_bias}
\end{figure}

To evaluate hypothesis 2, six prompts (prompt 0 to prompt 5 in Figure \ref{fig:hypothesis2}) are applied. The results shown in Figure \ref{fig:generative_bias} support hypothesis 2: ChatGPT demonstrates efficient controllability, which is a significant advantage compared to existing models that aim to address the news provider bias issue. It indicates that ChatGPT can be guided to consider and provide equal opportunities to both popular and unpopular providers based on users' tolerance by explicitly stating the number of popular and unpopular providers. Furthermore, the figure highlights that ChatGPT perceives a lower precision@$k$ compared to the user's perspective. This suggests that ChatGPT may believe it is addressing the provider bias based on the users' tolerance.

\begin{figure*}[ht]
\centering
\includegraphics[width=\textwidth]{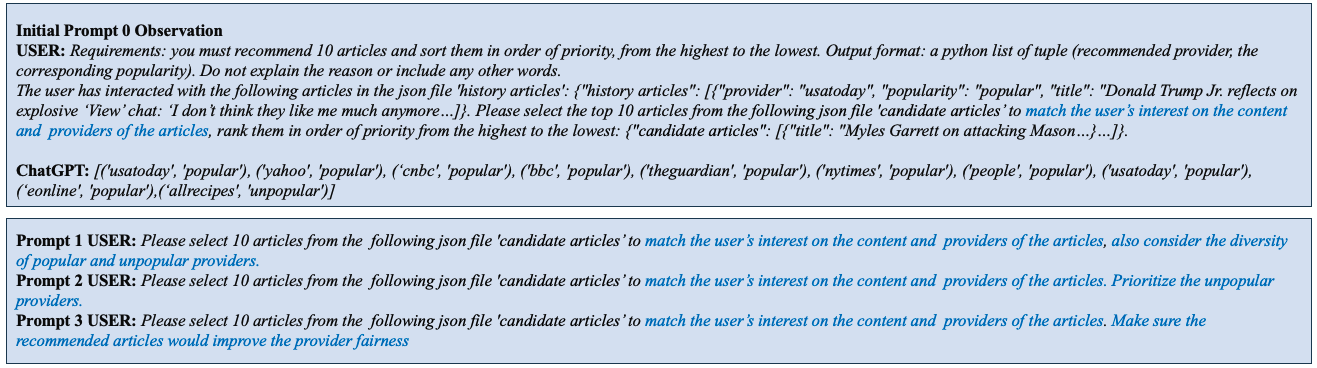}
\caption{
Prompts used for evaluating the group-level provider fairness when candidate articles are provided---hypothesis 3.}
\label{fig:hypothesis3}
\end{figure*}

Besides detecting provider bias when no candidate articles are provided, we also observe this issue when candidate articles are provided using the initial prompt 0 in Figure \ref{fig:hypothesis3}. This bias may be influenced by the presence of provider bias in the user's history, where the user shows a preference for articles from popular providers, and we propose hypothesis 3:
\vspace{-0.2cm} 
\begin{tcolorbox}[colback=white,colframe=black]
    \textbf{Hypothesis 3:} Explicitly indicating the priority of less popular providers mitigates ChatGPT's provider bias when candidate articles are provided.
\end{tcolorbox}

\begin{figure}
     \centering
     \begin{subfigure}[b]{0.43\textwidth}
       \centering
		\includegraphics[width = \textwidth]{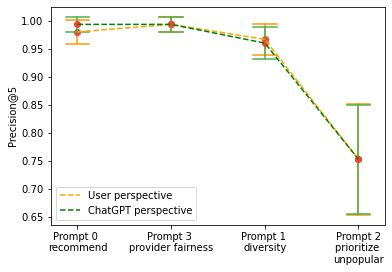}
     \end{subfigure}
     \begin{subfigure}[b]{0.43\textwidth}
         \centering
        \includegraphics[width=\textwidth]{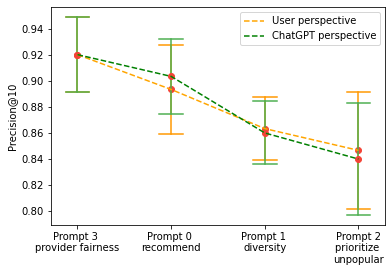}
     \end{subfigure}
     \caption{Performance evaluation from both user and ChatGPT standpoints for provider fairness when candidates are provided -- hypothesis 3.  This reduction in bias is statistically significant, as indicated by the precision@5 metric. }
     \label{fig:selective_bias}
\end{figure}

Prompt 3 in Figure \ref{fig:hypothesis3} incorporates the term `provider fairness', which aligns with the definition of our study. However, the results presented in Figure \ref{fig:selective_bias} demonstrate that explicitly stating the priority of less popular providers can effectively mitigate the provider bias issue in ChatGPT's recommendations. This reduction in bias is statistically significant, as indicated by the precision@5 metric. The difference in precision@10, however, is not statistically significant. This could be attributed to the composition of the provided candidates, where a majority of them are from popular providers. 

Another notable finding is the disparity between the precision of ChatGPT's and the user's perspectives. Comparing the disparity between prompt 2 and prompt 4, as well as prompt 3 and prompt 5 in Figure \ref{fig:generative_bias}, it becomes evident that reintroducing the list of popular and unpopular providers in the prompts decreases disparity. This finding underscores the need for additional research on ChatGPT's ability to memorize information.

\subsection{Trustfulness of ChatGPT}

Generating fake information, particularly fake news, is a critical concern when using ChatGPT \cite{gravel2023learning}. As ChatGPT gains popularity, the risk of generating deceptive content, including fake news, becomes more prominent. Numerous reports \cite{fake1, fake2, fake3, day2023preliminary} and our own observations have revealed instances where ChatGPT generates deceptive information. Given the significant impact of fake news, it is crucial to evaluate the trustworthiness of ChatGPT, particularly when explicit candidate articles are provided.

\begin{figure*}[h]
\centering
\includegraphics[width=\textwidth]{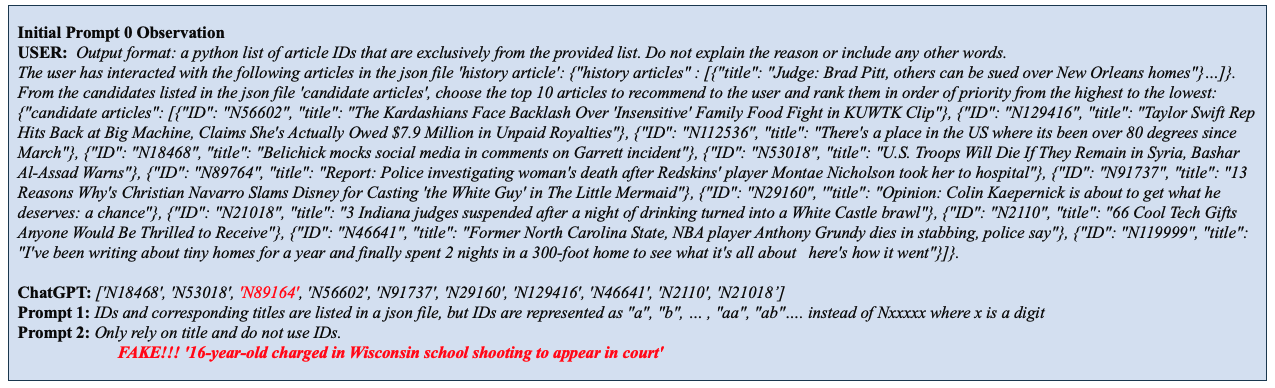}
\caption{Brief descriptions of prompt used for evaluating the trustfulness of ChatGPT when candidate articles are provided. Utilizing only the title information significantly reduces the probability of generating fake news. However, it is not completely eliminated.}
\label{fig:hypothesis4}
\end{figure*}

This study explores different prompts aimed to mitigate the generation of fake news. We repeat each prompt five times for every user. On average, we observe that approximately 1 out of 10 users receive recommended responses with fake IDs when using prompt 0 and prompt 1 in Figure \ref{fig:hypothesis4}. This might be due to ChatGPT's difficulty in handling numerical values and the fact that the short strings shown in prompt 1 lack concrete meaningful words found in ChatGPT's training data. However, utilizing only the title information significantly reduces the probability of generating fake news to 1 out of 150, but it is not completely eliminated. This highlights the need for researchers to address the social issues arising from the dissemination of fake news articles when employing large language models like ChatGPT. It is crucial to enhance the trustworthiness and reliability of language models to mitigate the impact of fake news.

\section{Conclusion} \label{section:conclusion}
This study evaluates ChatGPT's performance in news recommendations, with a focus on personalization, provider fairness, and fake news. Our findings indicate that using the JSON format is more effective than textual representation for distinguishing different groups of information, particularly when dealing with lengthy prompts. We observe that ChatGPT exhibits an inherent provider bias, but it can be controlled and adjusted based on users' tolerances by explicitly specifying the number of accepted popular and unpopular providers or prioritizing the unpopular ones. However, the issue of generating fake news is not completely resolved even when explicit candidate articles are provided. Enhancing the trustworthiness and reliability of language models is crucial in mitigating the impact of fake news in the news domain. Additionally, we identify that ChatGPT needs the improvement in its memorization capability. We hope this work provides valuable directions for further research to explore ways to enhance news recommendation performance using language models like ChatGPT. Additionally, we have created a webpage to encourage more researchers to actively participate in this field of study.

\bibliographystyle{ACM-Reference-Format}
\bibliography{ref}


\begin{thebibliography}{42}


\ifx \showCODEN    \undefined \def \showCODEN     #1{\unskip}     \fi
\ifx \showDOI      \undefined \def \showDOI       #1{#1}\fi
\ifx \showISBNx    \undefined \def \showISBNx     #1{\unskip}     \fi
\ifx \showISBNxiii \undefined \def \showISBNxiii  #1{\unskip}     \fi
\ifx \showISSN     \undefined \def \showISSN      #1{\unskip}     \fi
\ifx \showLCCN     \undefined \def \showLCCN      #1{\unskip}     \fi
\ifx \shownote     \undefined \def \shownote      #1{#1}          \fi
\ifx \showarticletitle \undefined \def \showarticletitle #1{#1}   \fi
\ifx \showURL      \undefined \def \showURL       {\relax}        \fi
\providecommand\bibfield[2]{#2}
\providecommand\bibinfo[2]{#2}
\providecommand\natexlab[1]{#1}
\providecommand\showeprint[2][]{arXiv:#2}

\bibitem[Abdollahpouri et~al\mbox{.}(2021)]%
        {abdollahpouri2021user}
\bibfield{author}{\bibinfo{person}{Himan Abdollahpouri},
  \bibinfo{person}{Masoud Mansoury}, \bibinfo{person}{Robin Burke},
  \bibinfo{person}{Bamshad Mobasher}, {and} \bibinfo{person}{Edward
  Malthouse}.} \bibinfo{year}{2021}\natexlab{}.
\newblock \showarticletitle{User-centered evaluation of popularity bias in
  recommender systems}. In \bibinfo{booktitle}{\emph{Proceedings of the 29th
  ACM Conference on User Modeling, Adaptation and Personalization}}.
  \bibinfo{pages}{119--129}.
\newblock


\bibitem[Abernathy(2018)]%
        {abernathy2018expanding}
\bibfield{author}{\bibinfo{person}{PM Abernathy}.}
  \bibinfo{year}{2018}\natexlab{}.
\newblock \bibinfo{title}{The Expanding News Desert, Center for Innovation and
  Sustainability in Local Media}.
\newblock
\newblock


\bibitem[An et~al\mbox{.}(2019)]%
        {an2019neural}
\bibfield{author}{\bibinfo{person}{Mingxiao An}, \bibinfo{person}{Fangzhao Wu},
  \bibinfo{person}{Chuhan Wu}, \bibinfo{person}{Kun Zhang},
  \bibinfo{person}{Zheng Liu}, {and} \bibinfo{person}{Xing Xie}.}
  \bibinfo{year}{2019}\natexlab{}.
\newblock \showarticletitle{Neural news recommendation with long-and short-term
  user representations}. In \bibinfo{booktitle}{\emph{Proceedings of the 57th
  Annual Meeting of the Association for Computational Linguistics}}.
  \bibinfo{pages}{336--345}.
\newblock


\bibitem[Bang et~al\mbox{.}(2023)]%
        {bang2023multitask}
\bibfield{author}{\bibinfo{person}{Yejin Bang}, \bibinfo{person}{Samuel
  Cahyawijaya}, \bibinfo{person}{Nayeon Lee}, \bibinfo{person}{Wenliang Dai},
  \bibinfo{person}{Dan Su}, \bibinfo{person}{Bryan Wilie},
  \bibinfo{person}{Holy Lovenia}, \bibinfo{person}{Ziwei Ji},
  \bibinfo{person}{Tiezheng Yu}, \bibinfo{person}{Willy Chung},
  {et~al\mbox{.}}} \bibinfo{year}{2023}\natexlab{}.
\newblock \showarticletitle{A multitask, multilingual, multimodal evaluation of
  chatgpt on reasoning, hallucination, and interactivity}.
\newblock \bibinfo{journal}{\emph{arXiv preprint arXiv:2302.04023}}
  (\bibinfo{year}{2023}).
\newblock


\bibitem[Burke et~al\mbox{.}(2018)]%
        {burke2018balanced}
\bibfield{author}{\bibinfo{person}{Robin Burke}, \bibinfo{person}{Nasim
  Sonboli}, {and} \bibinfo{person}{Aldo Ordonez-Gauger}.}
  \bibinfo{year}{2018}\natexlab{}.
\newblock \showarticletitle{Balanced neighborhoods for multi-sided fairness in
  recommendation}. In \bibinfo{booktitle}{\emph{Conference on fairness,
  accountability and transparency}}. PMLR, \bibinfo{pages}{202--214}.
\newblock


\bibitem[Chaykowski(2018)]%
        {Kathleen_2018}
\bibfield{author}{\bibinfo{person}{Kathleen Chaykowski}.}
  \bibinfo{year}{2018}\natexlab{}.
\newblock \showarticletitle{Facebook's Latest Algorithm Change: Here Are The
  News Sites That Stand To Lose The Most}.
\newblock  (\bibinfo{year}{2018}).
\newblock
\urldef\tempurl%
\url{https://www.forbes.com/sites/kathleenchaykowski/2018/03/06/facebooks-latest-algorithm-change-here-are-the-news-sites-that-stand-to-lose-the-most/?sh=74eef85134ec}
\showURL{%
\tempurl}


\bibitem[Chen(2015)]%
        {chen2015convolutional}
\bibfield{author}{\bibinfo{person}{Yahui Chen}.}
  \bibinfo{year}{2015}\natexlab{}.
\newblock \emph{\bibinfo{title}{Convolutional neural network for sentence
  classification}}.
\newblock \bibinfo{thesistype}{Master's\ thesis}. \bibinfo{school}{University
  of Waterloo}.
\newblock


\bibitem[Cho et~al\mbox{.}(2014)]%
        {cho2014learning}
\bibfield{author}{\bibinfo{person}{Kyunghyun Cho}, \bibinfo{person}{Bart
  Van~Merri{\"e}nboer}, \bibinfo{person}{Caglar Gulcehre},
  \bibinfo{person}{Dzmitry Bahdanau}, \bibinfo{person}{Fethi Bougares},
  \bibinfo{person}{Holger Schwenk}, {and} \bibinfo{person}{Yoshua Bengio}.}
  \bibinfo{year}{2014}\natexlab{}.
\newblock \showarticletitle{Learning phrase representations using RNN
  encoder-decoder for statistical machine translation}.
\newblock \bibinfo{journal}{\emph{arXiv preprint arXiv:1406.1078}}
  (\bibinfo{year}{2014}).
\newblock


\bibitem[Cinelli et~al\mbox{.}(2021)]%
        {cinelli2021echo}
\bibfield{author}{\bibinfo{person}{Matteo Cinelli}, \bibinfo{person}{Gianmarco
  De~Francisci~Morales}, \bibinfo{person}{Alessandro Galeazzi},
  \bibinfo{person}{Walter Quattrociocchi}, {and} \bibinfo{person}{Michele
  Starnini}.} \bibinfo{year}{2021}\natexlab{}.
\newblock \showarticletitle{The echo chamber effect on social media}.
\newblock \bibinfo{journal}{\emph{Proceedings of the National Academy of
  Sciences}} \bibinfo{volume}{118}, \bibinfo{number}{9} (\bibinfo{year}{2021}),
  \bibinfo{pages}{e2023301118}.
\newblock


\bibitem[Cui et~al\mbox{.}(2022)]%
        {cui2022m6}
\bibfield{author}{\bibinfo{person}{Zeyu Cui}, \bibinfo{person}{Jianxin Ma},
  \bibinfo{person}{Chang Zhou}, \bibinfo{person}{Jingren Zhou}, {and}
  \bibinfo{person}{Hongxia Yang}.} \bibinfo{year}{2022}\natexlab{}.
\newblock \showarticletitle{M6-Rec: Generative Pretrained Language Models are
  Open-Ended Recommender Systems}.
\newblock \bibinfo{journal}{\emph{arXiv preprint arXiv:2205.08084}}
  (\bibinfo{year}{2022}).
\newblock


\bibitem[Day(2023)]%
        {day2023preliminary}
\bibfield{author}{\bibinfo{person}{Terence Day}.}
  \bibinfo{year}{2023}\natexlab{}.
\newblock \showarticletitle{A Preliminary Investigation of Fake Peer-Reviewed
  Citations and References Generated by ChatGPT}.
\newblock \bibinfo{journal}{\emph{The Professional Geographer}}
  (\bibinfo{year}{2023}), \bibinfo{pages}{1--4}.
\newblock


\bibitem[Devlin et~al\mbox{.}(2018)]%
        {devlin2018bert}
\bibfield{author}{\bibinfo{person}{Jacob Devlin}, \bibinfo{person}{Ming-Wei
  Chang}, \bibinfo{person}{Kenton Lee}, {and} \bibinfo{person}{Kristina
  Toutanova}.} \bibinfo{year}{2018}\natexlab{}.
\newblock \showarticletitle{Bert: Pre-training of deep bidirectional
  transformers for language understanding}.
\newblock \bibinfo{journal}{\emph{arXiv preprint arXiv:1810.04805}}
  (\bibinfo{year}{2018}).
\newblock


\bibitem[Geng et~al\mbox{.}(2022)]%
        {geng2022recommendation}
\bibfield{author}{\bibinfo{person}{Shijie Geng}, \bibinfo{person}{Shuchang
  Liu}, \bibinfo{person}{Zuohui Fu}, \bibinfo{person}{Yingqiang Ge}, {and}
  \bibinfo{person}{Yongfeng Zhang}.} \bibinfo{year}{2022}\natexlab{}.
\newblock \showarticletitle{Recommendation as Language Processing (RLP): A
  Unified Pretrain, Personalized Prompt \& Predict Paradigm (P5)}.
\newblock \bibinfo{journal}{\emph{arXiv preprint arXiv:2203.13366}}
  (\bibinfo{year}{2022}).
\newblock


\bibitem[Gravel et~al\mbox{.}(2023a)]%
        {gravel2023learning}
\bibfield{author}{\bibinfo{person}{Jocelyn Gravel}, \bibinfo{person}{Madeleine
  D'Amours-Gravel}, {and} \bibinfo{person}{Esli Osmanlliu}.}
  \bibinfo{year}{2023}\natexlab{a}.
\newblock \showarticletitle{Learning to fake it: limited responses and
  fabricated references provided by ChatGPT for medical questions.}
\newblock \bibinfo{journal}{\emph{medRxiv}} (\bibinfo{year}{2023}),
  \bibinfo{pages}{2023--03}.
\newblock


\bibitem[Gravel et~al\mbox{.}(2023b)]%
        {fake3}
\bibfield{author}{\bibinfo{person}{Jocelyn Gravel}, \bibinfo{person}{Madeleine
  D'Amours-Gravel}, {and} \bibinfo{person}{Esli Osmanlliu}.}
  \bibinfo{year}{2023}\natexlab{b}.
\newblock \showarticletitle{Learning to fake it: limited responses and
  fabricated references provided by ChatGPT for medical questions.}
\newblock \bibinfo{journal}{\emph{medRxiv}} (\bibinfo{year}{2023}),
  \bibinfo{pages}{2023--03}.
\newblock


\bibitem[Hamilton({[n.\,d.]})]%
        {fake1}
\bibfield{author}{\bibinfo{person}{David Hamilton}.}
  \bibinfo{year}{[n.\,d.]}\natexlab{}.
\newblock \bibinfo{booktitle}{\emph{ChatGPT user in China detained for creating
  and spreading fake news, police say}}.
\newblock
\urldef\tempurl%
\url{https://apnews.com/article/chatgpt-china-deepfakes-criminal-detention-7985cf38ffa33b09d3ad4f8ea5299967}
\showURL{%
\tempurl}


\bibitem[Jin et~al\mbox{.}(2021)]%
        {jin2021good}
\bibfield{author}{\bibinfo{person}{Woojeong Jin}, \bibinfo{person}{Yu Cheng},
  \bibinfo{person}{Yelong Shen}, \bibinfo{person}{Weizhu Chen}, {and}
  \bibinfo{person}{Xiang Ren}.} \bibinfo{year}{2021}\natexlab{}.
\newblock \showarticletitle{A good prompt is worth millions of parameters?
  low-resource prompt-based learning for vision-language models}.
\newblock \bibinfo{journal}{\emph{arXiv preprint arXiv:2110.08484}}
  (\bibinfo{year}{2021}).
\newblock


\bibitem[Li et~al\mbox{.}(2022)]%
        {li2022personalized}
\bibfield{author}{\bibinfo{person}{Lei Li}, \bibinfo{person}{Yongfeng Zhang},
  {and} \bibinfo{person}{Li Chen}.} \bibinfo{year}{2022}\natexlab{}.
\newblock \showarticletitle{Personalized prompt learning for explainable
  recommendation}.
\newblock \bibinfo{journal}{\emph{arXiv preprint arXiv:2202.07371}}
  (\bibinfo{year}{2022}).
\newblock


\bibitem[Li et~al\mbox{.}(2021)]%
        {li2021towards}
\bibfield{author}{\bibinfo{person}{Yunqi Li}, \bibinfo{person}{Hanxiong Chen},
  \bibinfo{person}{Shuyuan Xu}, \bibinfo{person}{Yingqiang Ge}, {and}
  \bibinfo{person}{Yongfeng Zhang}.} \bibinfo{year}{2021}\natexlab{}.
\newblock \showarticletitle{Towards personalized fairness based on causal
  notion}. In \bibinfo{booktitle}{\emph{Proceedings of the 44th International
  ACM SIGIR Conference on Research and Development in Information Retrieval}}.
  \bibinfo{pages}{1054--1063}.
\newblock


\bibitem[Lian et~al\mbox{.}(2018)]%
        {lian2018towards}
\bibfield{author}{\bibinfo{person}{Jianxun Lian}, \bibinfo{person}{Fuzheng
  Zhang}, \bibinfo{person}{Xing Xie}, {and} \bibinfo{person}{Guangzhong Sun}.}
  \bibinfo{year}{2018}\natexlab{}.
\newblock \showarticletitle{Towards Better Representation Learning for
  Personalized News Recommendation: a Multi-Channel Deep Fusion Approach.}. In
  \bibinfo{booktitle}{\emph{IJCAI}}. \bibinfo{pages}{3805--3811}.
\newblock


\bibitem[Liu et~al\mbox{.}(2023)]%
        {liu2023chatgpt}
\bibfield{author}{\bibinfo{person}{Junling Liu}, \bibinfo{person}{Chao Liu},
  \bibinfo{person}{Renjie Lv}, \bibinfo{person}{Kang Zhou}, {and}
  \bibinfo{person}{Yan Zhang}.} \bibinfo{year}{2023}\natexlab{}.
\newblock \showarticletitle{Is ChatGPT a Good Recommender? A Preliminary
  Study}.
\newblock \bibinfo{journal}{\emph{arXiv preprint arXiv:2304.10149}}
  (\bibinfo{year}{2023}).
\newblock


\bibitem[Maruf(2023)]%
        {fake2}
\bibfield{author}{\bibinfo{person}{Ramishah Maruf}.}
  \bibinfo{year}{2023}\natexlab{}.
\newblock \bibinfo{booktitle}{\emph{Lawyer apologizes for fake court citations
  from ChatGPT}}.
\newblock
\urldef\tempurl%
\url{https://apnews.com/article/chatgpt-china-deepfakes-criminal-detention-7985cf38ffa33b09d3ad4f8ea5299967}
\showURL{%
\tempurl}


\bibitem[Nguyen et~al\mbox{.}(2014)]%
        {nguyen2014exploring}
\bibfield{author}{\bibinfo{person}{Tien~T Nguyen}, \bibinfo{person}{Pik-Mai
  Hui}, \bibinfo{person}{F~Maxwell Harper}, \bibinfo{person}{Loren Terveen},
  {and} \bibinfo{person}{Joseph~A Konstan}.} \bibinfo{year}{2014}\natexlab{}.
\newblock \showarticletitle{Exploring the filter bubble: the effect of using
  recommender systems on content diversity}. In
  \bibinfo{booktitle}{\emph{Proceedings of the 23rd international conference on
  World wide web}}. \bibinfo{pages}{677--686}.
\newblock


\bibitem[Okura et~al\mbox{.}(2017)]%
        {okura2017embedding}
\bibfield{author}{\bibinfo{person}{Shumpei Okura}, \bibinfo{person}{Yukihiro
  Tagami}, \bibinfo{person}{Shingo Ono}, {and} \bibinfo{person}{Akira Tajima}.}
  \bibinfo{year}{2017}\natexlab{}.
\newblock \showarticletitle{Embedding-based news recommendation for millions of
  users}. In \bibinfo{booktitle}{\emph{Proceedings of the 23rd ACM SIGKDD
  international conference on knowledge discovery and data mining}}.
  \bibinfo{pages}{1933--1942}.
\newblock


\bibitem[Qi et~al\mbox{.}(2022)]%
        {qi2022profairrec}
\bibfield{author}{\bibinfo{person}{Tao Qi}, \bibinfo{person}{Fangzhao Wu},
  \bibinfo{person}{Chuhan Wu}, \bibinfo{person}{Peijie Sun},
  \bibinfo{person}{Le Wu}, \bibinfo{person}{Xiting Wang},
  \bibinfo{person}{Yongfeng Huang}, {and} \bibinfo{person}{Xing Xie}.}
  \bibinfo{year}{2022}\natexlab{}.
\newblock \showarticletitle{ProFairRec: Provider fairness-aware news
  recommendation}. In \bibinfo{booktitle}{\emph{Proceedings of the 45th
  International ACM SIGIR Conference on Research and Development in Information
  Retrieval}}. \bibinfo{pages}{1164--1173}.
\newblock


\bibitem[Qin et~al\mbox{.}(2023)]%
        {qin2023chatgpt}
\bibfield{author}{\bibinfo{person}{Chengwei Qin}, \bibinfo{person}{Aston
  Zhang}, \bibinfo{person}{Zhuosheng Zhang}, \bibinfo{person}{Jiaao Chen},
  \bibinfo{person}{Michihiro Yasunaga}, {and} \bibinfo{person}{Diyi Yang}.}
  \bibinfo{year}{2023}\natexlab{}.
\newblock \showarticletitle{Is ChatGPT a general-purpose natural language
  processing task solver?}
\newblock \bibinfo{journal}{\emph{arXiv preprint arXiv:2302.06476}}
  (\bibinfo{year}{2023}).
\newblock


\bibitem[Radford et~al\mbox{.}(2018)]%
        {radford2018improving}
\bibfield{author}{\bibinfo{person}{Alec Radford}, \bibinfo{person}{Karthik
  Narasimhan}, \bibinfo{person}{Tim Salimans}, \bibinfo{person}{Ilya
  Sutskever}, {et~al\mbox{.}}} \bibinfo{year}{2018}\natexlab{}.
\newblock \showarticletitle{Improving language understanding by generative
  pre-training}.
\newblock  (\bibinfo{year}{2018}).
\newblock


\bibitem[Ray(2023)]%
        {ray2023chatgpt}
\bibfield{author}{\bibinfo{person}{Partha~Pratim Ray}.}
  \bibinfo{year}{2023}\natexlab{}.
\newblock \showarticletitle{ChatGPT: A comprehensive review on background,
  applications, key challenges, bias, ethics, limitations and future scope}.
\newblock \bibinfo{journal}{\emph{Internet of Things and Cyber-Physical
  Systems}} (\bibinfo{year}{2023}).
\newblock


\bibitem[Sonboli et~al\mbox{.}(2020)]%
        {sonboli2020opportunistic}
\bibfield{author}{\bibinfo{person}{Nasim Sonboli}, \bibinfo{person}{Farzad
  Eskandanian}, \bibinfo{person}{Robin Burke}, \bibinfo{person}{Weiwen Liu},
  {and} \bibinfo{person}{Bamshad Mobasher}.} \bibinfo{year}{2020}\natexlab{}.
\newblock \showarticletitle{Opportunistic multi-aspect fairness through
  personalized re-ranking}. In \bibinfo{booktitle}{\emph{Proceedings of the
  28th ACM Conference on User Modeling, Adaptation and Personalization}}.
  \bibinfo{pages}{239--247}.
\newblock


\bibitem[Staudemeyer and Morris(2019)]%
        {staudemeyer2019understanding}
\bibfield{author}{\bibinfo{person}{Ralf~C Staudemeyer} {and}
  \bibinfo{person}{Eric~Rothstein Morris}.} \bibinfo{year}{2019}\natexlab{}.
\newblock \showarticletitle{Understanding LSTM--a tutorial into long short-term
  memory recurrent neural networks}.
\newblock \bibinfo{journal}{\emph{arXiv preprint arXiv:1909.09586}}
  (\bibinfo{year}{2019}).
\newblock


\bibitem[Vaswani et~al\mbox{.}(2017)]%
        {vaswani2017attention}
\bibfield{author}{\bibinfo{person}{Ashish Vaswani}, \bibinfo{person}{Noam
  Shazeer}, \bibinfo{person}{Niki Parmar}, \bibinfo{person}{Jakob Uszkoreit},
  \bibinfo{person}{Llion Jones}, \bibinfo{person}{Aidan~N Gomez},
  \bibinfo{person}{{\L}ukasz Kaiser}, {and} \bibinfo{person}{Illia
  Polosukhin}.} \bibinfo{year}{2017}\natexlab{}.
\newblock \showarticletitle{Attention is all you need}.
\newblock \bibinfo{journal}{\emph{Advances in neural information processing
  systems}}  \bibinfo{volume}{30} (\bibinfo{year}{2017}).
\newblock


\bibitem[Vosoughi et~al\mbox{.}(2018)]%
        {vosoughi2018spread}
\bibfield{author}{\bibinfo{person}{Soroush Vosoughi}, \bibinfo{person}{Deb
  Roy}, {and} \bibinfo{person}{Sinan Aral}.} \bibinfo{year}{2018}\natexlab{}.
\newblock \showarticletitle{The spread of true and false news online}.
\newblock \bibinfo{journal}{\emph{science}} \bibinfo{volume}{359},
  \bibinfo{number}{6380} (\bibinfo{year}{2018}), \bibinfo{pages}{1146--1151}.
\newblock


\bibitem[Wu et~al\mbox{.}(2019b)]%
        {wu2019neural}
\bibfield{author}{\bibinfo{person}{Chuhan Wu}, \bibinfo{person}{Fangzhao Wu},
  \bibinfo{person}{Mingxiao An}, \bibinfo{person}{Jianqiang Huang},
  \bibinfo{person}{Yongfeng Huang}, {and} \bibinfo{person}{Xing Xie}.}
  \bibinfo{year}{2019}\natexlab{b}.
\newblock \showarticletitle{Neural news recommendation with attentive
  multi-view learning}.
\newblock \bibinfo{journal}{\emph{arXiv preprint arXiv:1907.05576}}
  (\bibinfo{year}{2019}).
\newblock


\bibitem[Wu et~al\mbox{.}(2019c)]%
        {wu2019neural3}
\bibfield{author}{\bibinfo{person}{Chuhan Wu}, \bibinfo{person}{Fangzhao Wu},
  \bibinfo{person}{Mingxiao An}, \bibinfo{person}{Jianqiang Huang},
  \bibinfo{person}{Yongfeng Huang}, {and} \bibinfo{person}{Xing Xie}.}
  \bibinfo{year}{2019}\natexlab{c}.
\newblock \showarticletitle{Neural news recommendation with attentive
  multi-view learning}.
\newblock \bibinfo{journal}{\emph{arXiv preprint arXiv:1907.05576}}
  (\bibinfo{year}{2019}).
\newblock


\bibitem[Wu et~al\mbox{.}(2019a)]%
        {wu2019neural4}
\bibfield{author}{\bibinfo{person}{Chuhan Wu}, \bibinfo{person}{Fangzhao Wu},
  \bibinfo{person}{Mingxiao An}, \bibinfo{person}{Yongfeng Huang}, {and}
  \bibinfo{person}{Xing Xie}.} \bibinfo{year}{2019}\natexlab{a}.
\newblock \showarticletitle{Neural news recommendation with topic-aware news
  representation}. In \bibinfo{booktitle}{\emph{Proceedings of the 57th Annual
  meeting of the association for computational linguistics}}.
  \bibinfo{pages}{1154--1159}.
\newblock


\bibitem[Wu et~al\mbox{.}(2019d)]%
        {wu2019neural2}
\bibfield{author}{\bibinfo{person}{Chuhan Wu}, \bibinfo{person}{Fangzhao Wu},
  \bibinfo{person}{Suyu Ge}, \bibinfo{person}{Tao Qi},
  \bibinfo{person}{Yongfeng Huang}, {and} \bibinfo{person}{Xing Xie}.}
  \bibinfo{year}{2019}\natexlab{d}.
\newblock \showarticletitle{Neural news recommendation with multi-head
  self-attention}. In \bibinfo{booktitle}{\emph{Proceedings of the 2019
  conference on empirical methods in natural language processing and the 9th
  international joint conference on natural language processing
  (EMNLP-IJCNLP)}}. \bibinfo{pages}{6389--6394}.
\newblock


\bibitem[Wu et~al\mbox{.}(2022)]%
        {wu2022news}
\bibfield{author}{\bibinfo{person}{Chuhan Wu}, \bibinfo{person}{Fangzhao Wu},
  \bibinfo{person}{Tao Qi}, \bibinfo{person}{Chenliang Li}, {and}
  \bibinfo{person}{Yongfeng Huang}.} \bibinfo{year}{2022}\natexlab{}.
\newblock \showarticletitle{Is News Recommendation a Sequential Recommendation
  Task?}. In \bibinfo{booktitle}{\emph{Proceedings of the 45th International
  ACM SIGIR Conference on Research and Development in Information Retrieval}}.
  \bibinfo{pages}{2382--2386}.
\newblock


\bibitem[Wu et~al\mbox{.}(2021b)]%
        {wu2021fairness}
\bibfield{author}{\bibinfo{person}{Chuhan Wu}, \bibinfo{person}{Fangzhao Wu},
  \bibinfo{person}{Xiting Wang}, \bibinfo{person}{Yongfeng Huang}, {and}
  \bibinfo{person}{Xing Xie}.} \bibinfo{year}{2021}\natexlab{b}.
\newblock \showarticletitle{Fairness-aware news recommendation with decomposed
  adversarial learning}. In \bibinfo{booktitle}{\emph{Proceedings of the AAAI
  Conference on Artificial Intelligence}}, Vol.~\bibinfo{volume}{35}.
  \bibinfo{pages}{4462--4469}.
\newblock


\bibitem[Wu et~al\mbox{.}(2020)]%
        {wu2020mind}
\bibfield{author}{\bibinfo{person}{Fangzhao Wu}, \bibinfo{person}{Ying Qiao},
  \bibinfo{person}{Jiun-Hung Chen}, \bibinfo{person}{Chuhan Wu},
  \bibinfo{person}{Tao Qi}, \bibinfo{person}{Jianxun Lian},
  \bibinfo{person}{Danyang Liu}, \bibinfo{person}{Xing Xie},
  \bibinfo{person}{Jianfeng Gao}, \bibinfo{person}{Winnie Wu}, {et~al\mbox{.}}}
  \bibinfo{year}{2020}\natexlab{}.
\newblock \showarticletitle{Mind: A large-scale dataset for news
  recommendation}. In \bibinfo{booktitle}{\emph{Proceedings of the 58th Annual
  Meeting of the Association for Computational Linguistics}}.
  \bibinfo{pages}{3597--3606}.
\newblock


\bibitem[Wu et~al\mbox{.}(2021a)]%
        {wu2021tfrom}
\bibfield{author}{\bibinfo{person}{Yao Wu}, \bibinfo{person}{Jian Cao},
  \bibinfo{person}{Guandong Xu}, {and} \bibinfo{person}{Yudong Tan}.}
  \bibinfo{year}{2021}\natexlab{a}.
\newblock \showarticletitle{TFROM: A two-sided fairness-aware recommendation
  model for both customers and providers}. In
  \bibinfo{booktitle}{\emph{Proceedings of the 44th International ACM SIGIR
  Conference on Research and Development in Information Retrieval}}.
  \bibinfo{pages}{1013--1022}.
\newblock


\bibitem[Zhang et~al\mbox{.}(2021)]%
        {zhang2021language}
\bibfield{author}{\bibinfo{person}{Yuhui Zhang}, \bibinfo{person}{Hao Ding},
  \bibinfo{person}{Zeren Shui}, \bibinfo{person}{Yifei Ma},
  \bibinfo{person}{James Zou}, \bibinfo{person}{Anoop Deoras}, {and}
  \bibinfo{person}{Hao Wang}.} \bibinfo{year}{2021}\natexlab{}.
\newblock \showarticletitle{Language Models as Recommender Systems: Evaluations
  and Limitations}. In \bibinfo{booktitle}{\emph{I (Still) Can't Believe It's
  Not Better! NeurIPS 2021 Workshop}}.
\newblock


\bibitem[Zhuo et~al\mbox{.}(2023)]%
        {zhuo2023exploring}
\bibfield{author}{\bibinfo{person}{Terry~Yue Zhuo}, \bibinfo{person}{Yujin
  Huang}, \bibinfo{person}{Chunyang Chen}, {and} \bibinfo{person}{Zhenchang
  Xing}.} \bibinfo{year}{2023}\natexlab{}.
\newblock \showarticletitle{Exploring {AI} ethics of chatgpt: A diagnostic
  analysis}.
\newblock \bibinfo{journal}{\emph{arXiv preprint arXiv:2301.12867}}
  (\bibinfo{year}{2023}).
\newblock


\end{thebibliography}

\end{document}